# The composite picture of the charge carriers in $La_{2-x}Sr_xCuO_4$ ($0.063 \leqslant x \leqslant 0.11$) superconductors

**Y H Kim[1], P H Hor[2,3], X L Dong[3,4], F Zhou[4], Z X Zhao[4], Y S Song[3] and W X Ti[4]**

[1] Department of Physics, University of Cincinnati, Cincinnati, OH 45221-0011, USA
[2] Department of Physics, University of Houston, Houston, TX 77204-5005, USA
[3] Texas Center for Superconductivity, University of Houston, Houston, TX 77204-5002, USA
[4] National Laboratory for Superconductivity, Institute of Physics and Centre for Condensed Matter Physics, Chinese Academy of Sciences, PO Box 603, Beijing 100080, People's Republic of China

**Abstract**
Through far-infrared studies of $La_{2-x}Sr_xCuO_4$ single crystals for $x = 0.063$, 0.07, 0.09, and 0.11, we found that only $\sim$0.2% of the total holes participated in the nearly dissipationless normal state charge transport and superconductivity. We have also observed characteristic collective modes at $\omega \sim 18$ and 22 cm$^{-1}$ due to the bound carriers in an electronic lattice (EL) state, and the free carriers are massively screened by the EL. Our findings lead us to propose a composite picture of the charge system where the free carriers are coupled to and riding on the EL. This unique composite system of charge carriers may provide further insights into the understanding of the cuprate physics.

(Some figures in this article are in colour only in the electronic version)

## 1. Introduction

The superconductivity with high transition temperature ($T_c$) in copper-oxide-based compounds (cuprates) is one of the most puzzling problems in physics. More than one and a half decades after the discovery of superconductivity in cuprates, while substantial progress has been made, the underlying physics, especially the mechanism that leads to the high $T_c$, remains highly controversial. Therefore, it is legitimate to question whether some fundamental building blocks on which we have built our understanding of cuprate physics are incomplete or simply missing. In particular, we have examined the validity of the overwhelming single-component treatment, both theoretically and experimentally, of the doping-induced excess carriers (hereafter holes) in the $CuO_2$ plane. In this report, we wish to establish this missing link by presenting clear experimental evidence for a unique composite picture of the charge system of cuprates.

In the past few years, based on the neutron scattering study of the La$_{2-x-y}$Nd$_y$Sr$_x$CuO$_4$ system [1], charge and spin stripes have emerged as a possible topology of the charge inhomogeneity in the CuO$_2$ planes of the superconducting La$_{2-x}$Sr$_x$CuO$_4$ (LSCO). More recently, based on the observation of the pinned Goldstone modes and the single particle excitation gap at $\sim$0.05 eV in the far-infrared (far-IR) studies of the cation and anion co-doped polycrystalline La$_{1.895}$Sr$_{0.015}$CuO$_{4+\delta}$ samples at the hole concentration $p = 0.063$ and 0.07, the square lattice order with p(4 $\times$ 4) and c(2 $\times$ 2) symmetries was proposed as a plausible electronic structure of the bound holes [2, 3]. Other experimental signatures indicating a square lattice ordering of the holes had been reported in an electron diffraction study of La$_{1.875}$Ba$_{0.125}$CuO$_4$ [4] and in a transmission electron microscopy study of the La$_2$CuO$_{4+\delta}$ system [5].

Most recently, the detailed ac susceptibility studies of the La$_{2-x}$Sr$_x$CuO$_{4+\delta}$ system under high pressure showed evidence for an intrinsic electronic phase separation of the holes into two 'electronic phases' that support the superconductivity at two discrete temperatures, one at $\sim$15 K and the other at $\sim$30 K [6]. An electronic phase separation into spin glass ($x \sim 0.02$) and insulating ($x = 0$) phases was also observed in the lightly doped ($x < 0.02$) non-superconducting LSCO single crystals [7]. Indeed, although the topology of the charge arrangements was not demonstrated, recent direct experimental evidence for charge inhomogeneities in the CuO$_2$ planes [8–10] seems to be consistent with the above observations. However, despite the credible presence of the electronic phases whose relevant energy scales are within the reach of the far-IR probe energies, there has been no self-consistent far-IR single-crystal work that clearly demonstrates the existence of such electronic phases. Only recently have some far-IR results interpreted as charge density waves (CDWs) [11] or one-dimensional charge stripes [12–14] been reported.

In this paper, we report our findings of the far-IR signatures of the composite charge system of the free and bound holes. Through the in-plane ($ab$-plane) far-IR studies of a series of LSCO single crystals, we have observed a screened plasma edge at the frequency ($\omega$) between 15 and 20 cm$^{-1}$, depending on the doping level. This local minimum in the $ab$-plane reflectivity comes from an extremely small free hole density ($n_F$) which is only $\sim$0.2% of the total holes. This small fraction of the holes is responsible for the nearly dissipationless normal state charge transport and superconductivity. We found that these free holes are massively screened by the rest of the holes condensed into electronic lattices (ELs) in the CuO$_2$ planes.

## 2. Experimental details

For the measurement of the $ab$-plane reflectivity we used high quality La$_{2-x}$Sr$_x$CuO$_4$ single crystals grown by the travelling-solvent floating-zone method with $x = 0.063$, 0.07, 0.09, and 0.11. The growth and characterizations of the high quality crystal have been reported elsewhere [15]. The samples were prepared from pure, subgrain-free large as-grown ingots. Double-crystal x-ray rocking-curve and Rutherford backscattering spectrometry (RBS) ion channelling measurements indicated an estimated crystal mosaicity of 0.1° and a minimum RBS-channelling yield of 3.8% for the $x = 0.09$ crystal [15]. Both the dc resistivity and Meissner effect measurements of all four crystals are shown in figures 1(a) and (b), respectively. The sample area was 5 mm in diameter and the angular error from the desired crystal axis was less than $\pm 1°$.

We measured the reflectivity ($R$) at a near normal angle of incidence ($\sim$8°) on the sample. A Bruker 113v spectrometer was used to cover frequencies ($\omega$) between 10 cm$^{-1}$ ($\sim$1.2 meV) and 4000 cm$^{-1}$ (0.5 eV) for the $ab$-plane reflectivity. As a reference, we used a gold (Au) mirror made by depositing Au film on a stainless steel mirror that had been polished under the same conditions as the single-crystal samples. We corrected the measured reflectivity in reference to the Au mirror by multiplying the absolute reflectivity of the Au film calculated from the absolute measurement of the surface resistance ($r_s$) of Au via $R = 1 - 4r_s$ [16]. In order to cover the $\omega$ below 30 cm$^{-1}$, we mounted a 75 $\mu$m Mylar beamsplitter and a composite Cu-doped Si bolometer with 1 cm$^2$ active area operating at 2 K in conjunction with a parabolic light cone with a 7 mm diameter exit aperture. We controlled the sample temperature by directly monitoring the temperature ($T$) from the backside of the sample. We calculated the real part of the conductivity $\sigma_1(\omega)$ and the real part of the dielectric function $\varepsilon_1(\omega)$ via a Kramers–Kronig transformation of the measured reflectivity. We found that the asymptotic behaviour of the normal state reflectivity below 12 cm$^{-1}$ closely followed the Hagen–Rubens approximation. For high $\omega$ extrapolation above 4000 cm$^{-1}$ (0.5 eV), we used the data published by Uchida *et al* [17] for LSCO single crystals, which are nearly temperature independent and cover $\omega$ up to 37 eV.

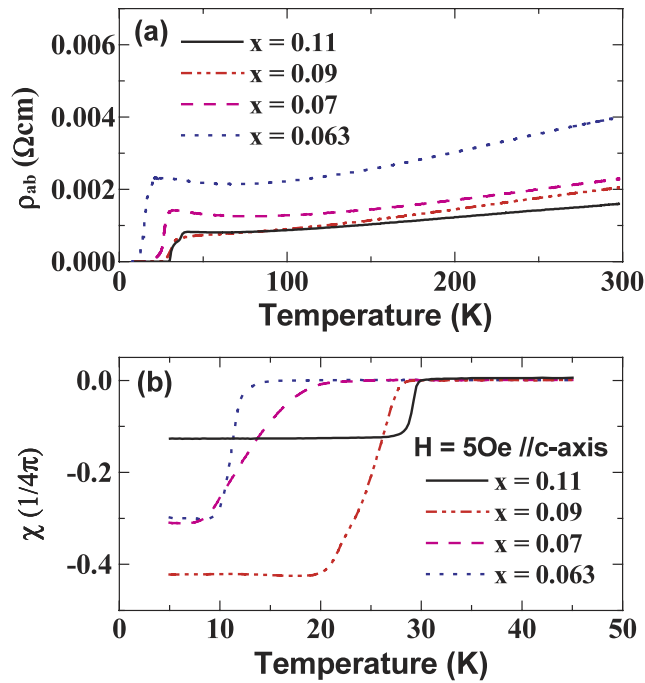

**Figure 1.** Plots of (a) the in-plane resisitivity and (b) the field cooled (5 Oe) measurements of La$_{2-x}$Sr$_x$CuO$_4$ single-crystal samples for $x = 0.063, 0.07, 0.09$ and $0.11$ with applied field along the $c$-axis.

## 3. Results and discussions

In figure 2, we present the unsmoothened far-IR *ab*-plane reflectivities (except for $\omega \lesssim 13$ cm$^{-1}$) of the four single-crystal samples at various $T$. Overall reflectivity is high for all samples. As $T$ decreases below 300 K, the $R$ reaches a maximum at $\omega \sim 150$ cm$^{-1}$ and then starts to slowly decrease until it reaches a local minimum at around $\omega \sim 20$ cm$^{-1}$ with decreasing $\omega$. Near the local reflectivity minimum, there is a sharp peak(s) which grows as $T$ decreases. The familiar phonons of the CuO$_2$ planes are readily seen in the reflectivity.

The corresponding Kramers–Kronig-derived $\sigma_1(\omega)$ plot in figure 3 shows highly unexpected charge dynamics. There develops an intense asymmetric structure located at around $\omega \sim 110$ cm$^{-1}$. This intense structure consists of a broad peak centred at $\omega \sim 60$ cm$^{-1}$ and another stronger peak at $\omega \sim 90$ cm$^{-1}$ as well as the *ab*-plane phonon modes [18]. In addition, a new mode ($\omega_{\mathrm{GL}}$) appears at $\omega \sim 18$ cm$^{-1}$ for $x = 0.063$ ($T_c \sim 16$ K). At low $T$ the $\omega_{\mathrm{GL}} \sim 18$ cm$^{-1}$ mode of the $x = 0.07$ ($T_c \sim 20$ K) sample takes over the 22 cm$^{-1}$ mode ($\omega_{\mathrm{GH}}$) seen at $T = 300$ K (see figure 4), and below $T_c$ a new mode ($\omega_{\phi\mathrm{L}}$) develops at $\sim 16$ cm$^{-1}$. For the $x = 0.09$ ($T_c \sim 28$ K) sample, the $\omega_{\mathrm{GH}}$ mode is more pronounced and the development of the new mode, $\sim 20$ cm$^{-1}$ ($\omega_{\phi\mathrm{H}}$), is evident at $T < T_c$. For $x = 0.11$ ($T_c \sim 30$ K), both the $\omega_{\mathrm{GL}}$ and $\omega_{\mathrm{GH}}$ modes are present and the emergence of the new modes denoted as $\omega_{\phi\mathrm{L}}$ and $\omega_{\phi\mathrm{H}}$ at $T < T_c$ becomes much clearer. The $\omega_{\phi\mathrm{L}}$ and $\omega_{\phi\mathrm{H}}$ modes that appear only when $T < T_c$ are obviously related to the development of the superconducting state. These modes are assigned as the superconducting phase collective modes due to the broken longitudinal gauge symmetry [2]. However, at the present time it is not clear whether the assigned $\omega_\phi$ for $x = 0.11$ are also present above $T_c$ because their positions appear to coincide with the ripple noise positions. The detailed discussion of the $T$-dependence of the oscillator strength of the $\omega_\phi$ and the $\omega_{\mathrm{G}}$ is beyond the scope of this paper and will be reported later.

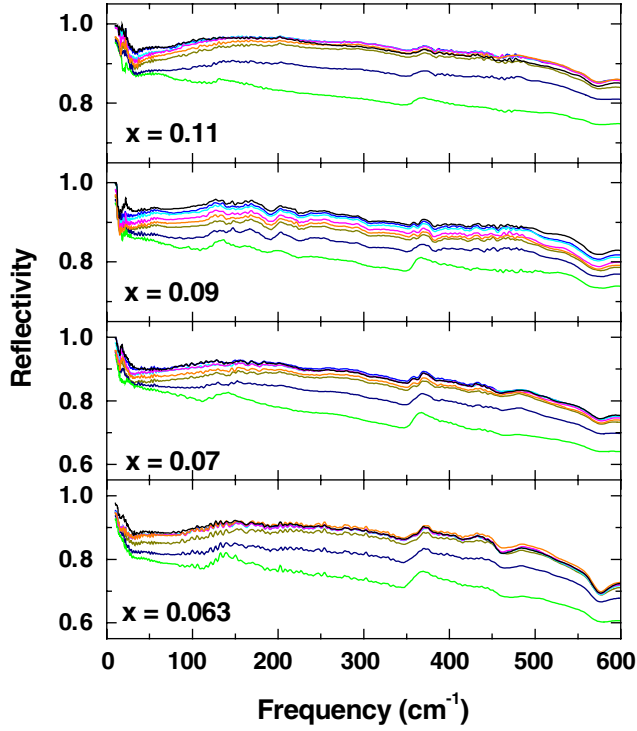

**Figure 2.** Reflectivity spectra for $x = 0.063$, 0.07, 0.09, and 0.11 at various temperatures. Top to bottom: $T = 8$ K (black), 16 K (blue), 24 K (cyan), 30 K (magenta), 50 K (orange), 100 K (dark green), 200 K (navy), and 300 K (green). Note the common presence of the local minimum and the development of sharp structures near $\sim 20$ cm$^{-1}$ for all the samples.

The $T = 300$ K $\sigma_1(\omega)$ and $\varepsilon_1(\omega)$ for the $\omega$ range between 10 and 100 cm$^{-1}$ are displayed in figure 4. From the absence of a structure in $\sigma_1(\omega)$ at the zero crossing in $\varepsilon_1(\omega)$, it is clear that the local reflectivity minimum is the screened plasma edge of the free carriers in the CuO$_2$ planes. Note the extremely narrow Drude-like peak in $\sigma_1(\omega)$ with a half-width at half-maximum $\Gamma \sim 10$ cm$^{-1}$ and the collective modes at $\omega \sim 18$ and $\sim 22$ cm$^{-1}$. Such a small $\Gamma$ implies that the free holes in the CuO$_2$ planes transport charges without much dissipation in the normal state. It is this portion of the carriers that condenses into the superfluid state as evidenced by the disappearance of this Drude-like peak at $T \leqslant T_c$ (see the insets in figure 3). From the measured dc conductivity and taking $\Gamma \sim 10$ cm$^{-1}$, one can estimate the plasma frequency $\omega_p$ from $\omega_p^2 = 60\sigma_{dc}\Gamma$ to find $\omega_p \sim 388, 511, 542$, and 612 cm$^{-1}$ for $x = 0.063, 0.07, 0.09$, and 0.11 single crystals respectively. Here $\sigma_{dc}$ is in $\Omega^{-1}$ cm$^{-1}$ and $\Gamma$ in cm$^{-1}$. Comparing the above unscreened $ab$-plane $\omega_p$ with the screened plasma frequency $\omega_p^* = \omega_p/\sqrt{\varepsilon_0}$ found from the zero crossing in $\varepsilon_1(\omega)$, $\omega_0$ via $\omega_p^* = \sqrt{\omega_0^2 + \Gamma^2}$, we estimate the corresponding room temperature dielectric constant $\varepsilon_0 \sim 560, 880, 856$, and 963 for $x = 0.063, 0.07, 0.09$, and 0.11 respectively. From the unscreened $\omega_p$, the known total hole concentrations, and using the free electron mass for the free holes, we find only 0.17%, 0.25%, 0.22%, and 0.24% of the holes are contributing to the charge transport at room temperature for $x = 0.063$, 0.07, 0.09, and 0.11 respectively.

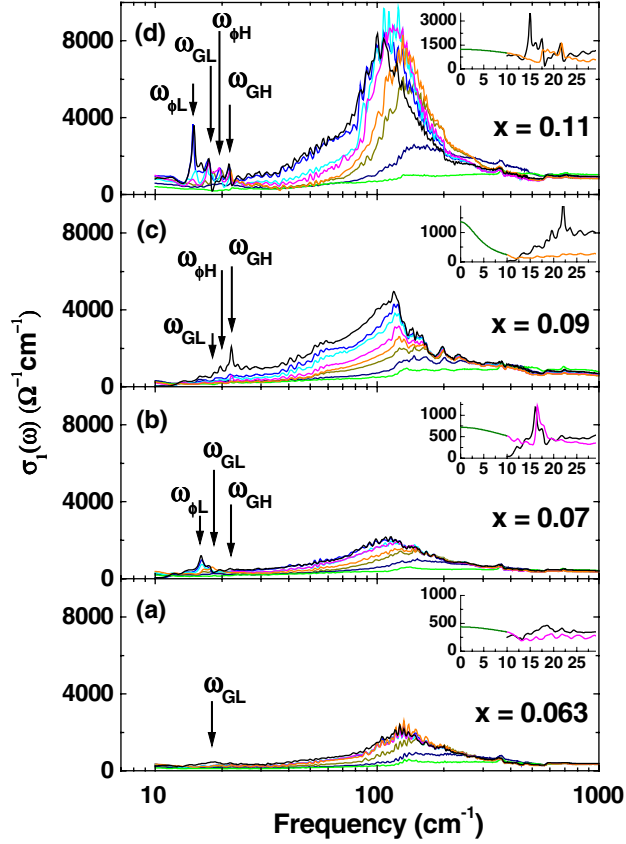

**Figure 3.** The corresponding frequency-dependent conductivity calculated from the reflectivity in figure 2. (a) For $x = 0.063$ ($T_c \sim 16$ K), there exists the collective mode ($\omega_{GL}$) at $\sim$18 cm$^{-1}$. At $T < T_c$, the 18 cm$^{-1}$ peak further increases its strength due to the presence of the phase collective mode ($\omega_{\phi L}$) at around 18 cm$^{-1}$. (b) In $x = 0.07$ ($T_c \sim 20$ K), the 22 cm$^{-1}$ mode ($\omega_{GH}$) is stronger than the 18 cm$^{-1}$ mode ($\omega_{GL}$) at room temperature (see figure 4). For $T < T_c$, the phase collective mode ($\omega_{\phi L}$) develops at $\sim$16 cm$^{-1}$. (c) For $x = 0.09$ ($T_c \sim 28$ K), the development of the phase collective mode near 20 cm$^{-1}$ ($\omega_{\phi H}$) is evident at $T < T_c$. (d) For $x = 0.11$ ($T_c \sim 30$ K), the separation among $\omega_{GL}$, $\omega_{\phi L}$, $\omega_{GH}$ and $\omega_{\phi H}$ become clearer (see the text). Insets: the Drude-like tail of the $\sigma_1(\omega)$ and its disappearance at $T < T_c$ (black, 8 K; magenta, 30 K; orange, 50 K). Olive curves below 10 cm$^{-1}$ in the $\sigma_1(\omega)$ plot are calculated from the Drude model using the measured dc conductivity.

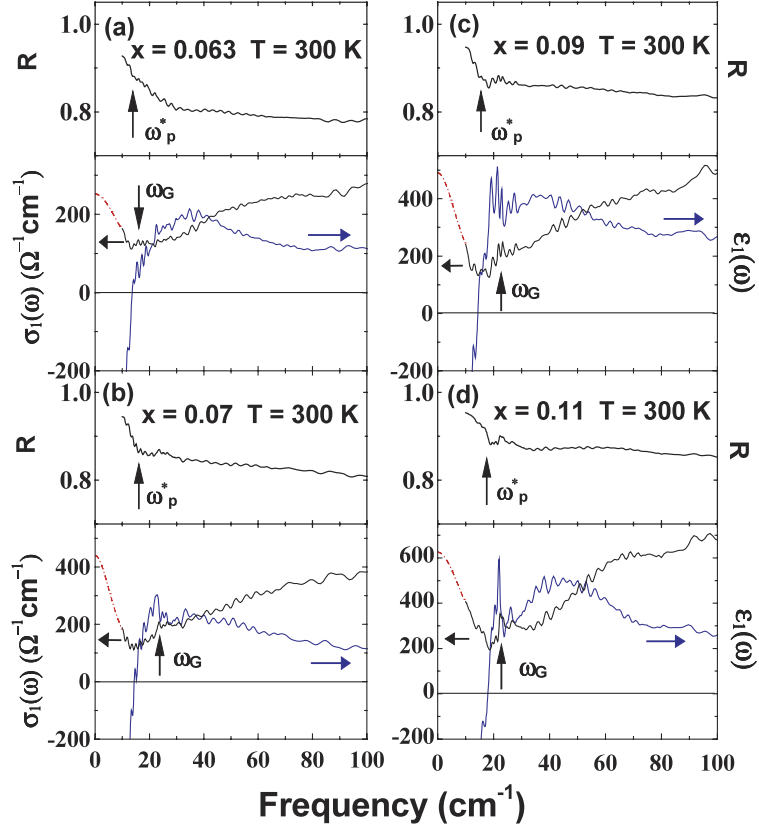

**Figure 4.** Detailed comparison of the room temperature reflectivity with the corresponding $\sigma_1(\omega)$ and $\varepsilon_1(\omega)$ (see the text). Red dash-dot curves below 10 cm$^{-1}$ in the $\sigma_1(\omega)$ plot are from the Drude fits. The Goldstone mode ($\omega_G$) and the screened plasma frequency ($\omega_p^*$) are indicated with an arrow. Note the same scale for both $\sigma_1(\omega)$ and $\varepsilon_1(\omega)$.

Since all the free carrier contribution to the reflectivity is confined to $\omega < 20$ cm$^{-1}$, it is clear that the high $ab$-plane reflectivity in the commonly accessed far-IR region (30–400 cm$^{-1}$) is not due to the metallic behaviour of the CuO$_2$ planes as usually assumed but due to the large imaginary part of the index of refraction arising from strong and rapidly varying absorption structures. Therefore, the measured reflectivity of cuprates is now sensitive to the angle of incidence and polarization of the far-IR [19]. This has not been a problem for normal Drude-like metallic systems where the displacement current of the bound charges is negligible [20, 21]. However, for the $\omega$-range where the displacement current is dominating, the absolute reflectivity of the sample can be higher than that of the metallic reference mirror. This difference is sufficient to make the apparent measured reflectivity in reference to a metallic mirror exceed unity as often seen in the alkali halide crystals in the region between the transverse and longitudinal phonon modes. At the 8° angle of incidence, we found no difference between the normal state far-IR reflectivity measured with the $\pi$ polarization and that measured with the $\sigma$ polarization.

We point out that the $\sigma_1(\omega)$ below 100 cm$^{-1}$ displayed in figure 3 do not resemble any of the previously published normal state far-IR data [12, 18, 22–26] even though the presence of the intense electronic structure at $\omega \sim 110$ cm$^{-1}$ is evident in their superconducting state $\sigma_1(\omega)$. The primary source of this problem lies in the experimental limitation on the angle of incidence when the reflectivity was measured. This is now particularly important for cuprates. For example, at the 15° angle of incidence [18], instead of the decreasing reflectivity from the maximum at $\omega \sim 110$ cm$^{-1}$ with decreasing $\omega$ as seen in this work, the reflectivity monotonically increases to ~0.95 at $\omega \sim 50$ cm$^{-1}$ as if it follows the free electron behaviour. In a case of 8° angle incidence [13, 14], a similar result as the present work was obtained except for the inconsistent $T$-dependence. More importantly, we did not observe the change in the position of the ~110 cm$^{-1}$ structure with doping as Lucarelli *et al* described [14, 27]. Therefore, without the correct reflectivity information below 50 cm$^{-1}$, the tail on the high frequency side of the intense structure peaked at $\omega \sim 110$ cm$^{-1}$ can easily be misidentified as the Drude-like tail because this structure also grows with doping [17, 18, 26].

During the course of this work we found that even increasing the angle of incidence by ~1° from our minimum 8° angle of incidence tends to amplify the low frequency (below 50 cm$^{-1}$ in particular) spectral structures in the reflectivity to give stronger absorption structures in $\sigma_1(\omega)$ with the portions of their peaks being negative, which is unphysical. Even at our 8° angle of incidence, the low $T$ reflectivities for $\omega \lesssim 15$ cm$^{-1}$ and at $\omega \sim 150$ cm$^{-1}$ exceeds unity by ~1–2%, forcing us to rescale the overall reflectivity by matching the far-IR tail below 12 cm$^{-1}$ with the calculated Hagen–Rubens behaviour using the measured dc conductivity. However, even though such scaling may decrease the magnitude of the background conductivity and the strength of the absorption peaks, all the spectral features were preserved.

In order to self-consistently explain the above experimental observations of the massively screened free holes (~0.2% of the total holes) in the presence of the collective modes ($\omega_G$), we conclude that the rest of the holes have to condense into an electronic lattice (EL) state to which the free carriers must be coupled. Furthermore, the free holes must 'ride' the EL to avoid the scattering with the phonons of the CuO$_2$ lattice. Hence, we expect the single-particle excitation gap ($2\Delta$) in the $\sigma_1(\omega)$ of the EL which is the condensation energy of the holes into the EL state and the $\omega_G$ which is the gapped Goldstone mode of the commensuration-pinned EL with the underlying CuO$_2$ lattice resulting from the strong hole–lattice interactions [28]. At the same time, the presence of $\omega_G$ gives $\varepsilon_1(\omega)$ of the EL as $\varepsilon_1^{EL}(\omega) = \varepsilon_{ab} + \Omega_{EL}^2/(\omega_G^2 - \omega^2) + \varepsilon_1(2\Delta)$ [29]. Here $\varepsilon_{ab}$ is the dielectric constant of the underlying CuO$_2$ lattice, $\varepsilon_1(2\Delta)$ is the single-particle excitation contribution, and $\Omega_{EL}$ is defined as $\Omega_{EL}^2 \equiv 4\pi n_{EL} e^2/m_{EL}$ ($n_{EL}$ = bound hole density and $m_{EL}$ = dynamic mass of the EL).

From this $\varepsilon_1^{EL}$, one may estimate the dynamic mass of the bound holes in the EL state. For instance, using $\varepsilon_1^{EL}(0) = \varepsilon_0 \sim 560$ and $\omega_G \sim 18$ cm$^{-1}$ and $n_{EL} \sim 6.3 \times 10^{20}$ holes cm$^{-3}$ (99.8% of the holes), we find $m_{EL} \sim 150$ $m_e$ ($m_e$ = free electron mass) for $x = 0.063$ at 300 K. This small $m_{EL}$ implies that the long range order of the EL has already been sufficiently developed at 300 K. For $x$ beyond 0.063, the mode at ~22 cm$^{-1}$ begins to dominate and develops more strongly as the doping increases (see figure 4). This new Goldstone mode must then be associated with a different EL state. The oscillator strength of the Goldstone mode continuously increases as $T$ decreases below 300 K and saturates at $T \sim 200$ K by reaching the dynamic mass ~90 $m_e$ (details are not shown), indicating that the development of the long range order of the ELs of $x = 0.063$ is completed at $T \sim 200$ K.

As a specific finger print, the dimensionality of the EL structure determines the shape of the single-particle excitation peak in $\sigma_1(\omega)$. If the EL has a one-dimensional structure, the square-root singularity in the joint density of states will produce a sharp, asymmetric peak at $\omega = 2\Delta$ [29]. If it is two dimensional, then the flat density of states will give rise to a steplike absorption structure at $\omega = 2\Delta$. In view of this, the $\sigma_1(\omega)$ shown in figure 3 does not appear to bear the signature of the one-dimensional EL. Therefore, we suggest that the EL in the LSCO is two dimensional in nature. Although the single-particle excitation energy gap is not clearly seen owing to the two-dimensional nature of the EL and the presence of the intense peak at $\omega \sim 110 \text{ cm}^{-1}$, we do not expect this gap to be different from that of the polycrystalline LSCO samples ($2\Delta \sim 0.05 \text{ eV}$) [2, 3].

It is interesting to note that, despite the high crystallographic order, the degree of electronic disorder in the single-crystalline LSCO samples is higher than that in the polycrystalline sample. For example when soft dopants [6], such as excess oxygen atoms that are highly mobile even below room temperature, were used in the Sr/O co-doped LSCO system, two distinct electronic superconducting phases with corresponding sharp Goldstone modes, one at $\omega_{GL} \sim 23 \text{ cm}^{-1}$ for the $T_c = 15 \text{ K}$ phase and the other at $\omega_{GH} \sim 46 \text{ cm}^{-1}$ for the $T_c = 30 \text{ K}$ phase, were observed [2]. Comparing the polycrystalline 7% Sr-doped LSCO sample with the Sr/O co-doped LSCO sample at $p = 0.07$ ($T_c \sim 26 \text{ K}$), it was found that the development of the $\omega_{GH} \sim 46 \text{ cm}^{-1}$ mode was suppressed in the absence of the soft dopants and the $T_c \sim 16 \text{ K}$ superconductivity results [3]. In the 7% Sr-doped LSCO single crystal, the corresponding Goldstone modes now appear at $\omega_{GL} \sim 18 \text{ cm}^{-1}$ and $\omega_{GH} \sim 22 \text{ cm}^{-1}$ (see figures 3 and 4) and the superconducting transition occurs at $T = 20 \text{ K}$. However, these modes seen in single-crystalline LSCO are much sharper than those observed in polycrystalline samples.

Judging from the lower frequencies and multiplicity of the Goldstone modes, we suggest that the ELs in single-crystalline LSCO are in a random disorder state due to the randomly dispersed hard dopants [4] that would inhibit the development of the preferred ELs. Therefore, a variety of lower order two-dimensional ELs may possibly be inter-woven in LSCO single crystals. The extremely broad structure developed at around $\omega \sim 40 \text{ cm}^{-1}$, which appears to be contributed by the modes at around $\omega \sim 30$ and $\sim 40 \text{ cm}^{-1}$ lumped together with the broad structure at $\omega \sim 60 \text{ cm}^{-1}$, reflects the influence of such electronic disorder. Furthermore, the smaller fraction of the free carriers in the 7% Sr-doped single-crystalline LSCO (0.25%) than that found in the corresponding polycrystalline sample (0.43%) suggests the presence of incomplete ELs caused by the more electronic disorder in the single-crystalline samples.

The physical origin of the absorption structure peaked at $\omega \sim 110 \text{ cm}^{-1}$ needs to be clarified. This mode results from the contributions of the broad peaks at $\omega \sim 60$ and $\sim 90 \text{ cm}^{-1}$ and of the phonon modes at $\omega \sim 100$ and $\sim 120 \text{ cm}^{-1}$. The relatively weak and sharp $\sim 120 \text{ cm}^{-1}$ mode, which is one of the well known $ab$-plane $E_u$ modes, is present on top of the strong doping-induced $E_g$ symmetry mode at $\sim 100 \text{ cm}^{-1}$ [18]. The oscillator strength of the doping-induced $E_g$ mode at $\omega \sim 100 \text{ cm}^{-1}$ increases with the doping and also with decreasing $T$. In the EL model, the broad peaks at $\omega \sim 60$ and $\sim 90 \text{ cm}^{-1}$ are due to the optical transition from the ground state of EL to the state generated by the free carriers coupled to the EL and another from the free hole state to the excited EL state [2, 3]. However, Dumm $et\ al$ [12] suggested the $\sim 110 \text{ cm}^{-1}$ mode as the structure due to the carrier localization peak while Lucarelli $et\ al$ [14] interpreted this mode as the collective mode of the charge stripes whose frequency depends on doping.

## 4. Summary and conclusion

Based on our detailed far-IR studies of the LSCO single crystals, the $ab$-plane charge dynamics can be understood as a composite system consisting of a small fraction of free holes that move on the EL formed by the rest of the holes. Our composite charge model provides a natural physical basis for some salient properties pertaining to the cuprates. For instance, assuming a square EL, this composite picture naturally explains the experimental observation that only $\sim 20\%$ of the holes participate in the superconducting condensate at optimum doping [3]. In the normal state, this small free carrier concentration can easily account for the observed unusual $c$-axis transport [30]. This charge model, in contrast to the recent theoretical discussions of the gossamer superconductivity within the resonating-valence-bond picture [31], also provides a natural experimental foundation for the gossamer nature of the high $T_c$ arising from the small superfluid density $n_F$.

We emphasize that the existence of free carriers and their transport are innately coupled to the underlying EL in this charge model unlike the common CDW system where the normal carriers are decoupled from the CDW condensate [32]. Our proposed charge model, if proven to be true, will place severe constraints on the microscopic model of the theory of high $T_c$ superconductivity and serve as one of the fundamental building blocks for the interpretation of all the previous experimental observations.

## Acknowledgments

We would like to acknowledge Mark Ankenbauer, Mark Sabatelli, and Bob Schrott for their fine machining work and Jiwu Xiong for her work in crystal growth and processing. PHH is supported by the State of Texas through the Texas Center for Superconductivity at the University of Houston. ZXZ, FZ, and WXT are supported by the Ministry of Science and Technology of China and the National Science Foundation of China through project G1999064601 and project 10174090.

## References

[1] Tranquada J M, Sternlieb B J, Axe J D, Nakamura Y and Uchida S 1995 *Nature* **375** 561
[2] Kim Y H and Hor P H 2001 *Mod. Phys. Lett.* B **15** 497
[3] Hor P H and Kim Y H 2002 *J. Phys.: Condens. Matter* **14** 10377
[4] Inoue Y, Wakabayashi Y, Ito K and Koyama Y 1994 *Physica* C **235–240** 835
[5] Li J Q, Chen L, Zhao Z X and Matsui Y 2000 *Physica* C **341–348** 1747
[6] Lorenz B, Li Z G, Honma T and Hor P H 2002 *Phys. Rev.* B **65** 144522
[7] Matsuda M, Fujita M, Yamada K, Birgeneau R J, Endoh Y and Shirane G 2002 *Phys. Rev.* B **65** 134515
[8] Iguchi I, Yamaguchi T and Sugimoto A 2001 *Nature* **412** 420
[9] Pan S H, O'Neal J P, Badzey R L, Chamon C, Ding H, Engelbrecht J R, Wang Z, Eisaki H, Uchida S, Guptak A K, Ng K W, Hudson E W, Lang K M and Davis J C 2001 *Nature* **413** 282
[10] Singer P M, Hunt A W and Imai T 2002 *Phys. Rev. Lett.* **88** 047602
[11] Bernhard C, Holden T, Humlicek J, Munzar D, Golnik A, Klaser M, Wolf T, Carr L, Homes C, Keimer B and Cardona M 2002 *Solid State Commun.* **121** 93
[12] Dumm M, Basov A N, Komiya S, Abe Y and Ando A 2002 *Phys. Rev. Lett.* **88** 147003
[13] Venturini F, Zhang Q M, Hackl R, Lucarelli A, Lupi S, Ortolani M, Calvani P, Kikugawa N and Fujita T 2002 *Phys. Rev.* B **66** 060502
[14] Lucarelli A, Lupi S, Ortolani M, Calvani P, Maselli P, Capizzi M, Giura P, Eisaki H, Kikugawa N, Fujita T, Fujita M and Yamada K 2003 *Phys. Rev. Lett.* **90** 037002
[15] Zhou F, Ti W X, Xiong J W, Zhao Z X, Dong X L, Hor P H, Zhang Z H and Chu W K 2003 *Supercond. Sci. Technol.* **16** L7
[16] Brändli G and Sievers A J 1972 *Phys. Rev.* B **5** 3550
[17] Uchida S, Ido T, Takagi H, Arima T, Tokura Y and Tajima S 1991 *Phys. Rev.* B **43** 7942
[18] Shimada M, Shimizu M, Tanaka J, Tanaka I and Kojima H 1992 *Physica* C **193** 277 and references therein
[19] See, for example, Klein M V 1970 *Optics* (New York: Wiley) chapter 11
[20] Reuter G E H and Sondheimer E H 1948 *Proc. R. Soc.* A **195** 336
[21] Dingle R B 1953 *Physica* **19** 311
[22] Timusk T, Puchkov A V, Basov D N and Startseva T 1998 *J. Phys. Chem. Solids* **59** 1953
[23] Basov D N, Liang R, Dabrowski B, Bonn D A, Hardy W N and Timusk T 1996 *Phys. Rev. Lett.* **77** 4090
[24] Wang N L, McConnell A W, Clayman B P and Gu G D 1999 *Phys. Rev.* B **59** 576
[25] Startseva T, Timusk T, Puchkov A V, Basov D N, Mook H A, Okuya M, Kimura T and Kishio K 1999 *Phys. Rev.* B **59** 7184

[26] Takenaka K, Shiozaki R, Okuyama S, Nohara J, Osuka A, Takayanagi Y and Sugai S 2002 *Phys. Rev.* B **65** 092405

[27] Tajima S, Uchida S, van der Marel D and Bosov D N 2003 *Phys. Rev. Lett.* **91** 129701
Lucarelli A, Lupi S, Ortolani M, Calvani P, Maselli P and Capizzi M 2003 *Phys. Rev. Lett.* **91** 129702

[28] Kim Y H, Heeger A J, Acedo L, Stucky G and Wudl F 1987 *Phys. Rev.* B **36** 7252

[29] Lee P A, Rice T M and Anderson P W 1974 *Solid State Commun.* **14** 703

[30] Kim Y H *et al* 2003 in preparation

[31] Zhang F C 2003 *Phys. Rev. Lett.* **90** 207002
Laughlin R B 2002 *Preprint* cond-mat/0209269
Yu Y 2002 *Preprint* cond-mat/0211131

[32] Beyermann W P, Mihály L and Grüner G 1986 *Phys. Rev. Lett.* **56** 1489